\documentclass[12pt,preprint]{aastex}

\begin{document}


\title{Detectability of Terrestrial Planets in Multi-Planet Systems: Preliminary Report}


\author{Wesley A. Traub\altaffilmark{1},
Charles Beichman\altaffilmark{2},
Andrew F. Boden\altaffilmark{2},
Alan P. Boss\altaffilmark{3},\\
Stefano Casertano\altaffilmark{4},
Joseph Catanzarite\altaffilmark{1},
Debra Fischer\altaffilmark{5},
Eric B. Ford\altaffilmark{6},
Andrew Gould\altaffilmark{7},
Sam Halverson\altaffilmark{8},
Andrew Howard\altaffilmark{8},
Shigeru Ida\altaffilmark{9},
N. Jeremy Kasdin\altaffilmark{10},\\
Gregory P. Laughlin\altaffilmark{11},
Harold F. Levison\altaffilmark{12},
Douglas Lin\altaffilmark{11},
Valeri Makarov\altaffilmark{2},
James Marr\altaffilmark{1},\\
Matthew Muterspaugh\altaffilmark{8,13},
Sean N. Raymond\altaffilmark{14},
Dmitry Savransky\altaffilmark{10},
Michael Shao\altaffilmark{1},\\
Alessandro Sozzetti\altaffilmark{15}, and
Cengxing Zhai\altaffilmark{1}}

\altaffiltext{1}{Jet Propulsion Laboratory, California Institute of Technology, Pasadena, CA 91109}
\altaffiltext{2}{NASA Exoplanet Science Institute, Pasadena, CA 91125}
\altaffiltext{3}{Carnegie Institution of Washington, Washington, DC 20015}
\altaffiltext{4}{Space Telescope Science Institute, Baltimore, MD 21218}
\altaffiltext{5}{San Francisco State University, San Francisco, CA 94132}
\altaffiltext{6}{University of Florida, Gainesville, FL 32611}
\altaffiltext{7}{Ohio State University, Columbus, OH 43210}
\altaffiltext{8}{University of California at Berkeley, Berkeley, CA 94720}
\altaffiltext{9}{Univ. of Tokyo, Tokyo, Japan}
\altaffiltext{10}{Princeton University, Princeton, NJ 08544}
\altaffiltext{11}{University of California at Santa Cruz, Santa Cruz, CA 95064}
\altaffiltext{12}{Southwest Research Institute, Boulder, CA 80302}
\altaffiltext{13}{Tennessee State Univ., Nashville, TN 37209}
\altaffiltext{14}{University of Colorado, Boulder, CO 80309}
\altaffiltext{15}{INAF-Osservatorio Astronomioco di Torino, Pino Torinese, Italy}





\begin{abstract}
We ask if Earth-like planets (terrestrial mass and habitable-zone orbit) can be detected in multi-planet systems, using astrometric and radial velocity observations.  We report here the preliminary results of double-blind calculations designed to answer this question.
\end{abstract}





\section{Introduction}

At first glance, it should not be difficult to extract the astrometric signal of an Earth-like planet from the composite signal of a system of planets around a star, because each planet has its own frequency in time, so in a Fourier analysis of the total signal, the signature of any given planet should stand out compared to any other planet.  However in reality the case might be not so simple, because, for example, a planet in an eccentric orbit with a dominating signal (e.g., a Jupiter) might have harmonic terms that are not recognized as such but might look like a separate planet, or a long-period planet observed over a time shorter than that period would have noise generated at many frequencies owing to the difficulty of distinguishing a proper motion on the sky from a part of an orbit.  For these reasons, plus the fact that a mission should always be simulated before it is flown, we initiated a double-blind simulation to see how well Earth-like planets (i.e., terrestrial masses, habitable-zone periods) in multi-planet systems could be detected with SIM Lite, with the help of radial velocity (RV) data.  An additional goal was to see what accuracy of SIM Lite is needed to detect Earth-like planets.

The simulation was organized with four teams of scientists, the planet modelers (Team A: PIs EF, GL, HL, DL, SR), the data simulators (Team B: AFB, VM), the data analyzers (Team C: PIs SC, DF, JK, MM, MS), and the overall summarizers (Team D: WT, CB, APB, AG, JM).

Team-A comprised five groups of planetary system modelers.  Each group generated about 150 planetary systems, using their own best estimate of the actual distribution of masses and periods in real systems.  The resulting systems were checked for nominal agreement with the observed statistic that about 10\% of planets are roughly Jupiter-like, to agree with the Cumming {\em et al.\/} (\cite{Cumming}) analysis of a Jupiter-complete sample of RV observations.  Each system is expected to be stable for at least 10 million years.

Team-B was a single group which took input planetary models, rotated the systems at random, set up realistic observing schedules, generated synthetic astrometric and RV signals, and added noise.

Team-C comprised five groups of data analyzers, competitively selected.  Each group was given the same practice data sets, with and without noise, to validate their code.  The groups worked independently to develop their own analysis codes.  The Team-C groups were asked to report planet signals detected in the joint astrometric-RV data streams that had a false-alarm probability of less than 1\%.  The groups were allowed 4 weeks to analyze all 48 systems.  The exercise was double-blind in the sense that the person distributing the simulated data to the analyzers did not know any details of the systems, so no hints could possibly be transmitted.

\section{Model planetary system parameters and simulations}

The reason for carrying out numerical simulations of the ability of SIM Lite to detect planets, instead of analytical calculations, is that the problem is highly nonlinear, and is therefore analytically intractable.  To carry out numerical simulations we must choose specific numerical values of the parameters.  For the task at hand, there are many potential parameters.  For the target star we can choose its mass, metallicity, distance, two ecliptic coordinates, its proper motion, its average radial velocity, the average and variance of star-spot coverage, the rotational period, the degree of astroseismic activity, and the total integration time per target.   For the astrometric reference stars we can choose the same parameters, plus the number, separation, and magnitude of each one.  For the planet we can choose the number of planets per system, the mass of each planet, seven orbital parameters per planet, and two spatial orientation parameters per system.   For the astrometric instrument we can choose the angular accuracy of locating the star on the sky on two axes, the degree to which the noise spectrum is gaussian or non-gaussian, the presence of a limiting noise floor, the photon collection efficiency, the slew speed between measurements, the timing of visits, the length of each visit, the target-reference star chopping strategy, the fraction of mission time devoted to exoplanet observations, the total length of the mission, and the space trajectory of the spacecraft.  For the RV observations we can choose the number of visits, the instrumental accuracy per visit, the astroseismic noise per visit, and other parameters similar to the astrometric instrument.  Finally, for the calculation itself, we can choose how many examples to run for each case in order to estimate a reliable answer, and to estimate the uncertainty of that answer.

Given the large number of dimensions to be explored, it is clear that a statistically significant calculation would require a prohibitively large amount of computing time.  Thus for the present task, we fixed some parameters to have representative values, typically the median value of a range, as described next.

\section{Fixed Parameters}

We chose to fix the following parameters, for the present task.  We fix the type of astrometric instrument to be the current best estimate (CBE) of the configuration and accuracy of SIM Lite.  (Note that ``accuracy'' means the deviation of an observation from the ``true'' value.  This is not the same as ``precision'', which refers to the smallness of the error.)  The SIM Lite instrument is configured as a spatial interferometer, conceptually similar to Albert Michelson's interferometer on the 100-inch telescope at Mt.\ Wilson in 1913, and to other ground-based amplitude interferometers since that time.  SIM Lite collects two patches of the incoming wavefront from a star, uses a mirror on a movable delay line to apply a suitable time delay to one of the wavefront patches, and combines the patches to form an interference fringe in the pupil plane.  The delay is adjusted to give the maximum visibility of the white-light interference fringe packet.  The origin of the delay line is defined as its position when the axis of the instrument is perpendicular to the propagation vector of the incoming wavefront, where the axis is the line between the pivot points of the two collecting mirrors at the ends of the instrument.  The distance between these pivots is the baseline of the instrument.  The sine of the angle between the instrument axis and the wavefront is the delay divided by the baseline. The accuracy of an angle measurement is determined by the total number of photo-electrons collected in the interference fringe pattern (mainly set by the brightness of the star, the collecting area, and the integration time), the shape of the pattern (mainly set by the length of the baseline and the wavelength) and knowledge of the baseline vector (direction in 3-space and length).

The relative position of a target star with respect to its group of reference stars is determined during a ``visit''. Each visit can be either short or long, depending on the measurement accuracy desired.  A ``short visit'' is about 2200 sec in length; a ``long visit'' is longer.   In a short visit, the target star's two-dimensional location in the plane of the sky is measured with respect to a local framework defined by the average location of 4 to 5 nearby reference stars.  The measurement along each axis takes half of the total, or 1100 sec.

During the 1100 sec time allocated to each axis in a short visit, the angle between the target star and the baseline vector is measured to a ``single measurement accuracy'' of about 1.0 $\mu$as.  Likewise the angle between each of the reference stars and the baseline vector is measured, and the results combined to give the average of the reference stars, to a similar single measurement accuracy of about 1.0 $\mu$as.  The angle between the target and reference group is the difference of these angles.  The uncertainty is the ``differential-measurement accuracy'', which is [(1.0 $\mu$as)$^2$ + (1.0 $\mu$as)$^2$]$^{1/2}$ = 1.4 $\mu$as along one axis. Likewise along the orthogonal axis.

Photon-counting statistics tells us that N measurements, each of $\sigma$ accuracy, can be combined to give a net accuracy of $\sigma_{N}$ = $\sigma$/N$^{1/2}$.  SIM Lite's laboratory measurements have shown that this rule works down to a noise floor of at least $\sigma_{floor}$ = 0.035 $\mu$as.  This says that up to N = ($\sigma_{N}$/$\sigma_{floor})^2$  = (1.4 $\mu$as/0.035 $\mu$as)$^2$ = 1600 such short measurements can be combined.

We fix the number of visits that SIM Lite makes to a given target star at exactly 250, over the mission lifetime.  This value is chosen to minimize the computational burden of this exercise; in a more realistic mission scenario, it could be variable. Combining 250 short visits gives an accuracy of $\sigma_{mission}$ = 1.4 $\mu$as/(250)$^{1/2}$ or about 0.089 $\mu$as.  Combining 250 long visits gives a better accuracy.  The length of each visit (2200 sec or greater), and therefore the uncertainty per visit, can be set according to the desired ultimate accuracy.

We fix the fraction of mission length devoted to integration on exoplanets at 40\%.  The time between visits is modeled here as a constant plus or minus a uniformly distributed random rectangular interval with an RMS that is 10\% of the mean time between observations.  We fix the sun exclusion angle (sun-spacecraft-target) for astrometry at 50 degrees.

In order to limit the number of parameters, for each particular example in the present task, the target star mass will be fixed at about 1.0 solar mass, the distance will be fixed at about 10 pc, and the ecliptic latitude will be fixed at about 30 deg.  In practice, a real SIM target star, with approximately these properties, was used.  Target and reference star-spot and astroseismology noise is included in the total noise.   Angular offsets between target and reference stars are reduced to a differential ecliptic coordinate system, so the explicit identity of the target star is not relevant for this task.

We fix the RV accuracy at 1 m/s RMS per measurement.  The length of the RV data stream is 15 years for a 5-year SIM Lite mission, and 20 years for a 10-year SIM Lite mission.  The rate of RV measurements is about 1 measurement per month.  The time between measurements is constant plus or minus a uniformly distributed random rectangular interval with an RMS that is 10\% of the mean time between observations. There will be times during the year when the target is not observable owing to the sun and the target's ecliptic latitude.  The sun exclusion angle for RV is 45 degrees.

The threshold criterion for detection of a planet is that the probability of detection is 50\% or larger.  As is shown in Scargle (\cite{Scargle}), to achieve a false alarm probability of about 1\%, within a factor of two, for a number of measurements that ranges from tens to thousands, and which therefore includes the likely number of measurements per target to be made by SIM Lite, the mission signal to noise ratio (SNR) needed is about 5.8, depending slightly on the number of measurements.  Thus a simple way of looking at the overall sensitivity of SIM is to say that if $\sigma$ is the one-axis RMS noise per differential measurement, N is the number of visits, and A is the RMS amplitude of the astrometric signature (after removal of any long-term drift) that we can detect with a probability of 50\%, then we have that
A = SNR $\times$ $\sigma$ /N$^{1/2}$         or         N = (SNR $\times$ $\sigma$/A )$^2$.

For example, the Earth amplitude at 10 pc is A = 0.3 $\mu$as, for a face-on orbit, so if the RMS measurement noise is $\sigma$= 1.4 $\mu$as, then to detect an Earth with SNR = 5.8, we need N = 733 measurements, comfortably within the noise-floor limit of 1600 measurements mentioned above.  As an illustration, at 1100 sec per measurement, or 2200 sec per visit, and 40\% of a 5-year mission available for on-star integration, this permits up to 39 stars to be observed, assuming all stars are equivalent to the Sun at 10 pc.  In practice, using the list of real stars, about 60 targets can be observed to Earth-like accuracy.

\section{Ignored Parameters}

The following astrophysical effects are not explicitly included in this phase of the study, because we anticipate that the noise from them is already included in the quoted uncertainties of measurement:  photon rate from reference stars, spots on target and reference stars, planets around reference stars, and astroseismic activity on target and reference stars.

The following astrophysical effects are explicitly ignored because we anticipate that they can be removed from the data with essentially perfect accuracy: relativistic effects in the orbit of the astrometric spacecraft, aberration of light, deflection of light by Jupiter and other bodies, and motion of the Earth as it affects the RV data (will be referenced to Solar System barycenter).

The following astrophysical effects are ignored in this task because although tractable, they introduce a greater degree of complexity in the data modeling and analysis than can be managed in a short time; we anticipate that they will be treated in the subsequent phase of the activity:  parallax of the reference stars, proper motion of the reference stars, and perspective acceleration of the target star.  For the purpose of this study, the reference stars will be considered at infinite distance, so that they are fixed in the sky, and their centroid is near the target star.  The apparent motion known as perspective acceleration is the change in parallax displacement resulting from variation in the distance of the star from the observer over time.
Since we desire that the target star has no perspective acceleration in the astrometric data sets, we specify that its radial velocity is zero at the mean observation time. At that instant, the star's velocity is transverse to the line of sight. However, since the systemic RV offset is an important parameter in the analysis of RV data, a random radial velocity offset is added to the reflex motion velocity in the RV data sets to maintain realism.  We also ignore slow instrumental drifts in the measurement of delay because we anticipate that these can be removed from the data.

\section{Varied Parameters}

Two types of planetary systems are considered, Solar-like systems, and theoretical model systems.

The Solar-like system planet parameters are the same as the present Solar System, but with each parameter perturbed by a rectangular distribution function that has an RMS value of 10\% of that parameter.  This allows a simulated test of the ability to detect an Earth in the presence of Jupiter, Saturn, etc., but without knowing the exact answer.  About 20\% of the examples are of this type.  The mean orbital plane of each of these is random with respect to the orbital plane of the astrometric instrument.  The ``blind spot'' mentioned in the ExoPTF report occurs when the exact distance to the target star (i.e., its parallax) is not known (always the case), and is therefore treated as an adjustable parameter, so that part of the reflex motion may become degenerate with its apparent motion owing to the orbit of the spacecraft.  To ensure that only about half of such situations result in a failed detection, any planetary system that has a planet with a period that is in the range from 0.90 to 1.10 year is rejected by Team B.

The theoretically-modeled systems will be selected from the ensemble of 500 or more systems generated by Team-A groups.  These groups were instructed to produce planet system models that are as close to our present knowledge as is possible, taking into account current observing biases.  About 80\% of the simulated systems are of this type.  We compared the statistical properties of these models with current knowledge, as represented by Cumming {\em et al.\/} (\cite{Cumming}), and found a sufficient degree of agreement to warrant saying that the ensemble of input systems is roughly similar to observations in this respect. For all planetary systems, the two orientation angles are randomized, and the times of periastron are randomized.

Two types of number of visits are considered, depending on the planet-mass detection goal.  One type is determined by the goal of detecting a 1 Earth-mass planet, and the other by a 3-Earth-mass planet, both at about 1 AU.    Thus the number of visits N is a variable parameter, depending on the desired final accuracy, however N will never exceed its upper limit of 1600, derived from the estimated upper limit to the noise floor.  These detection levels were calculated for single-planet systems, ignoring the potential extra noise that might result from a mission length that is shorter than the period of a massive and detectable long-period planet.  The purpose of this formulation is to answer the question of what accuracy is needed to detect Earths.

The two types of astrometric mission length are 5 and 10 years.   This explores a reasonable range of possibilities, but preserves the option for a 10-year mission in which the astrometric instrument will be able to see a closed orbit of a Jupiter-type planet, thus increasing the accuracy with which a terrestrial-mass planet can be found.  The reason for this is that it is possible to interpret a partial-orbit astrometric signature as the sum of a proper motion and a full orbit at shorter period.  Thus a 5-year mission might generate false terrestrial planets, but a 10-year mission more accurately distinguish between the two.

The total number of variable parameters is then 2 types of systems (Solar and random), 2 types of detection goals (1 and 3 Earths), and 2 types of mission length (5 and 10 years).  This gives 2$\times$2$\times$2 = 8 categories that must be considered.  Since the Team-C groups had only about a month to carry out their analyses, we limited the number of cases to be solved to 48 systems.  This means that about 6 examples of each category were calculated.   The number of examples per category is therefore just at the beginning edge of being statistically significant.  We assigned the same data sets to each Team C group, which gave us a good sample of how different algorithms perform on the identical data sets.  This gives another almost statistically significant measure of the efficacy of independent algorithms.  Overall, we believe that this balancing of number of parameters allowed us to answer the questions we set for ourselves.

\section{Preliminary Results}

A total of 48 planetary systems were generated, of which 32 were random systems from Team A, 8 were Solar-system analogs (perturbed), 4 were single terrestrial HZ planets, and 4 had no planets.  To focus on the key variables of planet mass and period, all simulations were for a single star at a fixed point in the sky and at 10 pc.  The RV noise was 1 m/s rms, a value that includes expected instrumental as well as astrophysical noise.  The astrometric noise for most of the data sets was the expected noise from SIM Lite, 1.0 $\mu$as per single 2200-sec observation per axis per star, and therefore a factor of 1.4  larger for a differential measurement (target with respect to reference) per axis (RA or dec).  The timelines for half of the data were 5 years of astrometric plus  15 years of RV observations, and for the other half 10 years of astro plus  20 years of RV.      The orbits were calculated using  independent Keplerian motion, i.e., n-body codes were not used.

Reliability of detection is the ratio of correct detections to the total of correct plus false alarms.  The reliability ranged from about 40\% to 100\%, with 3 groups being over 80\% (one group was not able to finish the exercise on time).  In principle, this value should have been about 99\%, if the false alarm rate had truly been 1\%.  However the short amount of time for the exercise meant that only one group (the one with prior experience) was able to fully weed out false alarms.  For this reason the exercise is being repeated with extra statistical tests added.

Completeness is the detected fraction of planets.  The completeness is expected to be poor (near 0) if the signal-to-noise ratio (SNR) is small compared to about 5.8, and it should be large (near 1) if the SNR is well above that value.  Here the SNR is defined as the rms amplitude of the true signal, divided by the noise for the entire observing campaign, which is the measurement noise per visit to the star divided by the square root of the number of visits.  This definition applies to astrometric as well as RV observations.  Over a range of SNR values from about 0.7 to 7000, we found that the completeness did indeed jump sharply from about 0 to 1, at an SNR of about 5.8, as expected theoretically.

The accuracy of results also was close to the theoretically expected value, for the key parameters of period and mass.  We calculated the expected accuracy using a minimum-variance bound method (Gould \cite{Gould}).  Comparing the orbital solutions from the Team-C groups with the actual model parameters, and scaling the differences by the expected uncertainty, we found a distribution with a roughly Gaussian core and broad wings.  Beyond the 3 sigma points, where essentially no data should fall, we found a handful of cases (14\%) which appear to be situations where the expected error was very small (say less than 1\% in mass or period), but the actual error was more than 3 times that value, in other words a good measurement but not as perfect as theoretically expected.

In overall summary, this first phase of the simulation showed that the answers to our initial questions are (1) yes, Earths can be detected in multi-planet systems, and (2) the sensitivity needed is almost exactly the posited situation of a 5-year SIM Lite mission, using 40\% of the available observing time, with the expected noise level and a 6-m baseline, plus the additional help (mostly with long-period planets) from 15 years of RV observations.  This first phase of study is being followed up with phase 2 in which all tentative detections will be subject to an additional statistical F-test and a stability test, as well as additional time for the analysis, before being declared as detections.

In conclusion, we note that the exercise generated great enthusiasm among the participants, each of whom put in an enormous amount of personal time to find the planets, and we learned a number of useful lessons along the way.

\acknowledgments
MM and AH are partially supported by the Townes Fellowship program at the UC Berkeley Space Sciences Laboratory.  We gratefully acknowledge conversations with Dimitri Veras and Scott Tremaine.  EBF acknowledges the University of Florida High-Performance Computing Center, and HL acknowledges the JPL Supercomputer Facility.  
Part of this research was carried out at the Jet Propulsion Laboratory, California Institute of Technology, under a contract with the National Aeronautics and Space Administration.

\clearpage


\end{document}